\documentclass[%
 reprint,
 amsmath,amssymb,
 aps,
prl
]{revtex4-2}

\usepackage{graphicx}% Include figure files
\usepackage{dcolumn}% Align table columns on decimal point
\usepackage{bm}% bold math
\begin{document}

\title{Negative Refraction of Terahertz Phonons via Interfacial Momentum Compensation}% Force line breaks with \\

\author{Hao Chen}
\author{Zhong-Ke Ding}
\author{Yuan Yao}
\author{Chang-Hao Ding}
\author{Nannan Luo}
\author{Jiang Zeng}
\author{Li-Ming Tang}
\email{Contact author: lmtang@semi.ac.cn}
\author{Ke-Qiu Chen}%
\email{Contact author: keqiuchen@hnu.edu.cn}
\affiliation{Department of Applied Physics, School of Physics and Electronics, Hunan University, Changsha 410082, China}

\date{\today}% It is always \today, today,

\begin{abstract}
Negative refraction provides a route to steer and focus wave energy flow, but it remains difficult to realize for coherent terahertz (THz) phonons. The difficulty stems from conventional dispersion-based mechanisms, which require strongly anisotropic or negative-curvature dispersions, while the long-wavelength acoustic phonons most favorable for coherent transport are nearly isotropic. Here we overcome this limitation by introducing a momentum compensation mechanism mediated by discrete translational symmetry. Discrete translational symmetry parallel to the interface supplies a compensating tangential momentum, reopening transmitted channels beyond the conventional critical condition and enabling negative refraction when this compensation reverses the tangential component. Mode-resolved calculations for hBN/graphene heterostructures establish this mechanism in laterally stitched in-plane interfaces and show how twisted van der Waals moir\'e superlattices shift the negative-refraction window to lower frequencies. These results identify periodic crystalline interfaces as symmetry-engineered elements for THz phonon momentum conversion and wavefront control.
\end{abstract}
\maketitle

%\tableofcontents
\emph{Introduction---}Controlling the direction of phonon energy flow is a central goal of phononics~\cite{RevModPhys.84.1045}, especially in the terahertz~(THz) regime where heat-carrying phonons are crystal Bloch waves rather than continuum elastic excitations~\cite{maldovan2013sound}. Negative refraction would provide a route to bend and focus such energy flow~\cite{pendry2000negative,PhysRevLett.92.127401,lee2015observation,hu2023gate}, but its realization for long-range coherent THz phonons remains difficult. Here the difficulty refers specifically to intrinsic lattice phonons, distinct from THz electromagnetic waves~\cite{PhysRevLett.102.023901,ling2018broadband}, van der Waals polaritons~\cite{hu2023gate,sternbach2023negative}, and topological wave systems~\cite{lange2024negative,liu2022all}, where negative refraction has been realized or proposed through effective electromagnetic parameters, strongly anisotropic polariton dispersions, or surface-arc states. At sonic and ultrasonic frequencies, phononic crystals~\cite{yang2004focusing,lu2007negative,he2018topological,wang2020tunable} and metamaterials~\cite{wu2011elastic,christensen2012anisotropic,garcia2014negative,xie2014wavefront,zanotto2022metamaterial} have achieved negative refraction by engineering band structures, isofrequency contours, and effective parameters. Extending this capability to THz phonons in crystals, however, is not a matter of simply miniaturizing these mesoscopic designs, because the relevant dispersion and momentum-matching conditions are constrained by the atomic lattice rather than freely engineered unit cells.

Conventional dispersion-based routes to negative refraction rely on strongly anisotropic~\cite{hu2023gate,christensen2012anisotropic,garcia2014negative,yang2004focusing,sternbach2023negative} or negative-curvature dispersions~\cite{feng2006acoustic,lu2007negative,PhysRevB.86.024301}. In crystals, such features typically occur in high-frequency optical branches or acoustic modes near the Brillouin-zone boundary. These modes are susceptible to interface-induced phase randomization~\cite{feng2015spectral,hu2018randomness} and anharmonic scattering~\cite{luckyanova2012coherent,luckyanova2018phonon}, limiting coherent propagation. By contrast, low-frequency acoustic phonons near the Brillouin-zone center dominate heat transport~\cite{pan2023ab,ding2024robustness,yao2026reflection,fhqs-mvfw} and retain coherence over much longer distances~\cite{maldovan2015phonon,qian2021phonon,zhang2021coherent,chen2021non}, but remain nearly isotropic, keeping the group velocity closely aligned with the wave vector. This mismatch has largely precluded a viable route to negative refraction for coherent, heat-carrying THz phonons.

Here we show that this bottleneck can be overcome at periodic crystalline interfaces, where the discrete translational symmetry preserved parallel to the boundary permits tangential momentum compensation set by the interfacial superperiod. This momentum transfer reopens transmitted channels beyond the conventional critical condition and enables negative refraction even for nearly isotropic THz acoustic phonons, without relying on strongly anisotropic bulk dispersions. Using hBN/graphene heterostructures and nonequilibrium Green's function calculations, we establish this mechanism in laterally stitched in-plane interfaces and extend it to twisted van der Waals heterostructures, where moir\'e superlattices provide a twist-tunable realization of the required interfacial periodicity.

\begin{figure}[t]
	\includegraphics{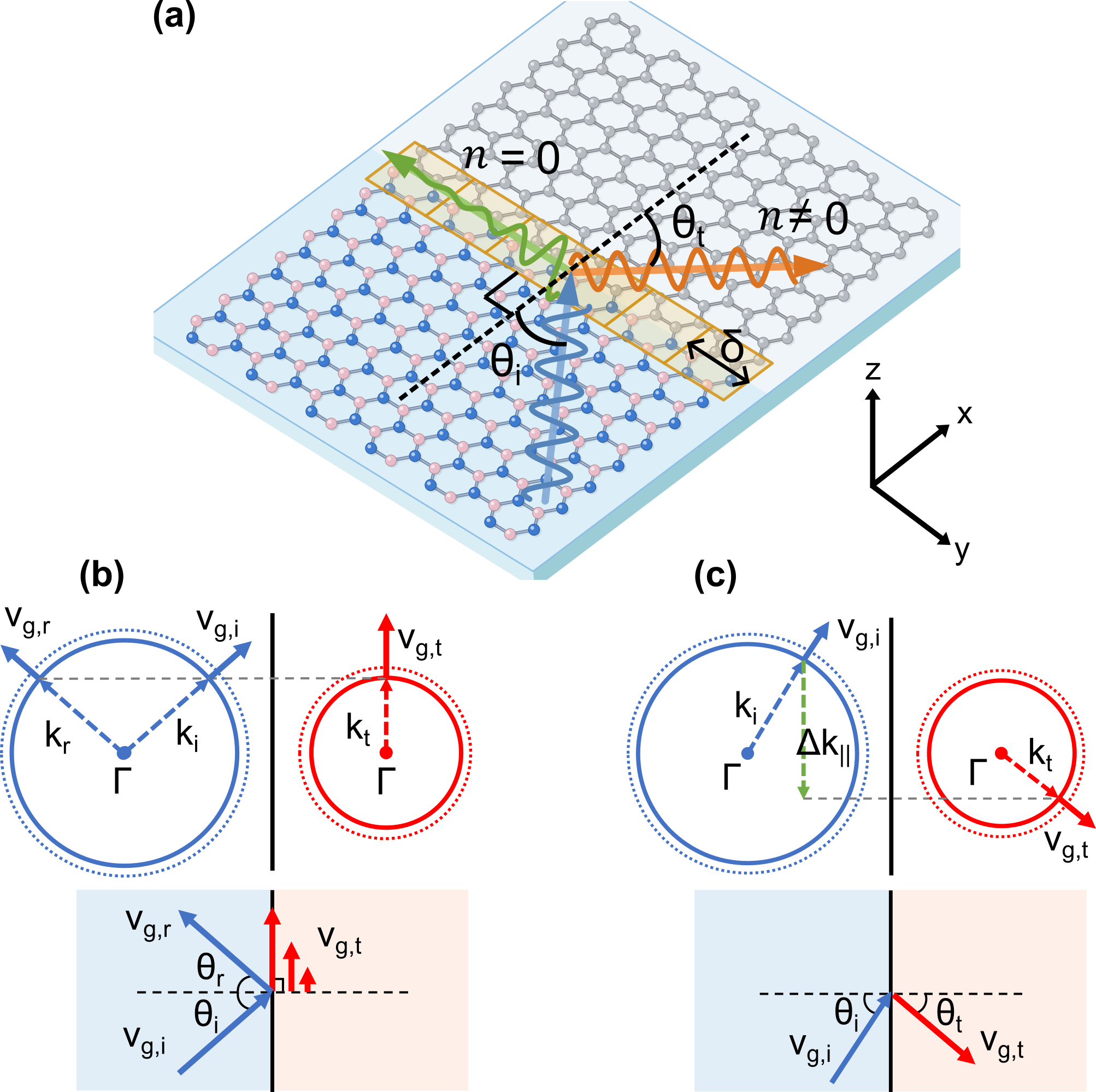}% Here is how to import EPS art
    \caption{
	Momentum-compensated negative refraction at a periodic crystalline interface.
	(a) Direct \(n=0\) and momentum-compensated \(n\neq 0\) transmission at an interface of period \(\delta\).
	(b) Strict conservation of \(k_{\parallel}\) yields an evanescent transmitted state beyond the conventional critical condition.
	(c) The shift \(\Delta k_{\parallel}=nG_{\parallel}\), with \(G_{\parallel}=2\pi/\delta\), reopens a propagating state with reversed tangential group velocity, producing negative refraction.
    }
	\label{fig:mechanism}
\end{figure}

\emph{Mechanism---}The key question is how negative refraction can occur when the acoustic isofrequency contours are nearly isotropic. For a laterally continuous interface, tangential crystal momentum is conserved, and the directly matched transmitted state leaves the propagating sector once the incidence angle exceeds the conventional critical angle~\cite{chen2022interfacial}. Its normal wave-vector component then becomes complex and the transmitted field is evanescent [Fig.~\ref{fig:mechanism}(b)]. A periodic crystalline interface changes this condition by allowing a discrete tangential momentum shift, \(\Delta k_\parallel=nG_\parallel\), where \(n\in\mathbb{Z}\) and \(G_\parallel=2\pi/\delta\). The \(n=0\) channel is the ordinary transmitted channel, whereas \(n\neq0\) channels are momentum-compensated channels that can remain propagating even after the direct channel has closed [Fig.~\ref{fig:mechanism}(a)]. Reopening alone, however, is not sufficient for negative refraction. For the low-frequency acoustic branches considered here, the group velocity remains closely aligned with the wave vector; a reopened channel becomes a negative-refraction channel only when the compensated tangential wave vector reverses sign relative to that of the incident phonon [Fig.~\ref{fig:mechanism}(c)].

This selection rule follows from the discrete translational symmetry preserved parallel to the interface. At the interface, scattering does not preserve the separate translational symmetries of the two bulk crystals, but only the subgroup that remains periodic along the boundary. Tangential crystal momentum is therefore conserved only modulo the interfacial reciprocal vector \(G_\parallel\). Writing \(k_\parallel\) as the signed tangential component, the transmitted channel in diffraction order \(n\) obeys the generalized Snell relation for phonons~\cite{yu2011light},
\begin{equation}
	k_{t,\parallel}^{(n)} = k_{i,\parallel}+nG_\parallel .
	\label{eq:snell}
\end{equation}
Equation~(\ref{eq:snell}) expresses the interfacial momentum-compensation mechanism. The interface supplies or absorbs a discrete tangential momentum \(nG_\parallel\), allowing a transmitted state forbidden under strict \(n=0\) matching to reenter the propagating sector. A more general derivation including an imposed phase gradient is given in the Supplemental Material~\cite{SuppM}. For the THz acoustic phonons considered here, the relevant contribution is the discrete \(nG_\parallel\) term generated by the crystalline interface itself~\cite{lu2025single}.

This mechanism leads to three predictions tested below. First, finite transmission should appear beyond the conventional critical boundary, where the continuous-interface picture predicts only evanescence. Second, after reopening, the compensated channel should enter the negative-refraction sector only if its tangential component changes sign. Third, the effect has a well-defined frequency window. For nearly isotropic acoustic phonons, let \(q_i(\omega)\) and \(q_t(\omega)\) denote the radii of the incident and transmitted isofrequency contours. For a channel that is closed under strict \(n=0\) matching, reopening requires the shifted tangential component \(k_{i,\parallel}+nG_\parallel\) to lie within the transmitted contour. Negative refraction further requires this shifted component to have the opposite sign to \(k_{i,\parallel}\). Combining these two requirements gives the existence condition for a momentum-compensated negative-refraction channel in the conventional total-reflection sector,
\begin{equation}
	q_t(\omega) \le |nG_\parallel| \le q_i(\omega)+q_t(\omega).
	\label{eq:window}
\end{equation}
The upper bound is the reopening condition, since the incident and transmitted contours together can bridge at most a tangential mismatch \(q_i+q_t\). The lower bound is the sign-reversal condition, since the interfacial momentum shift must be large enough to push the compensated transmitted state across \(k_\parallel=0\). For the nearly isotropic acoustic modes considered here, this \(k_{\parallel}\)-sign reversal corresponds to reversal of the tangential group-velocity component, while the incidence and refraction angles are determined directly from the group velocities of the lead modes. A detailed derivation and its application to the moir\'e twist sequence are given in the Supplemental Material. For the dominant first-order channel, the onset is therefore determined by
\begin{equation}
	q_i\!\left(\omega_{\rm on}^{(1)}\right)+q_t\!\left(\omega_{\rm on}^{(1)}\right)=G_\parallel ,
	\label{eq:onset}
\end{equation}
while the upper edge follows from \(q_t(\omega_{\rm off}^{(1)})=G_\parallel\). Increasing the interfacial period reduces \(G_\parallel\) and shifts the first-order negative-refraction window to lower frequencies.

\begin{figure}
	\includegraphics{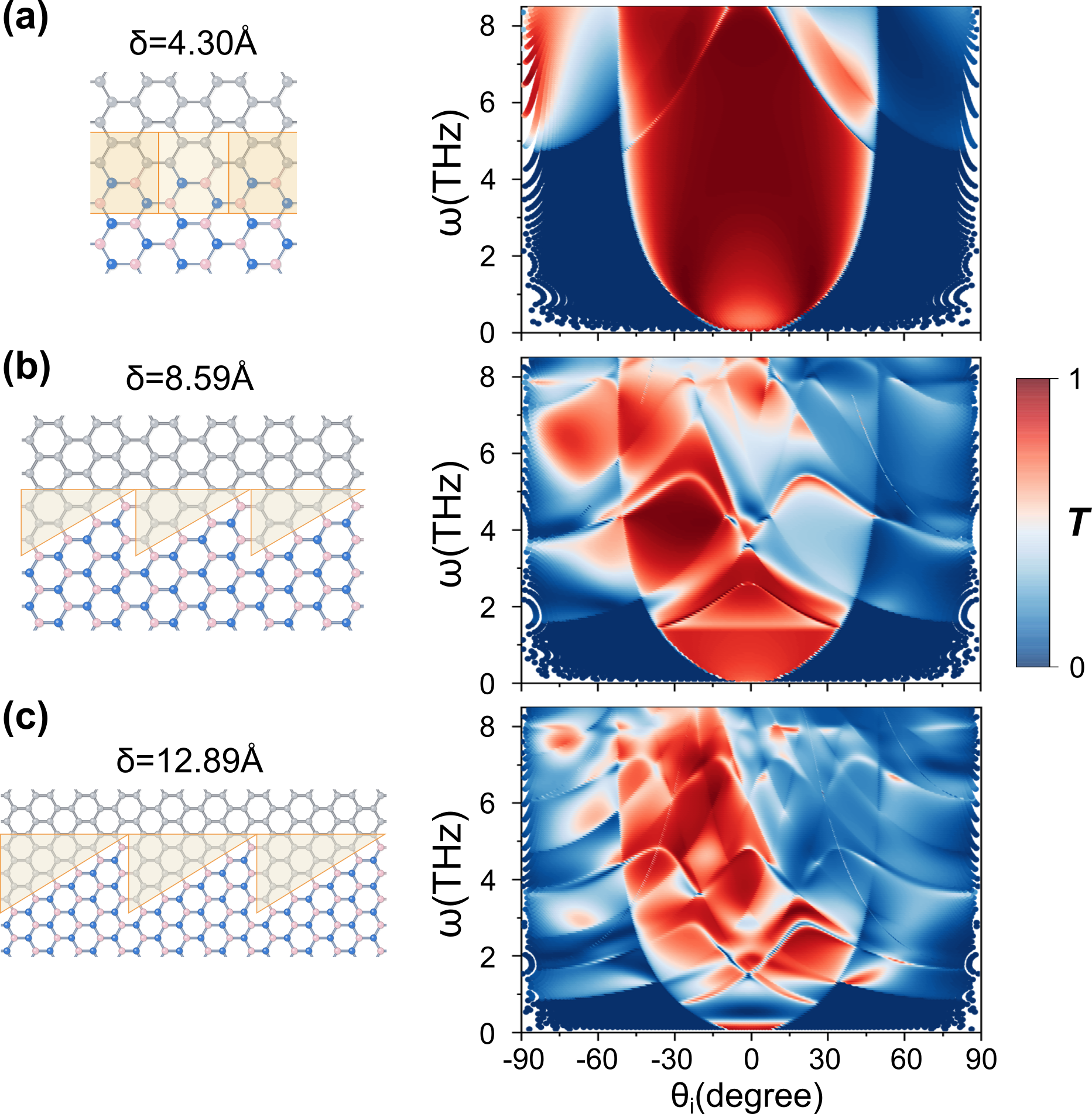}% Here is how to import EPS art
    \caption{
	Finite transmission beyond the conventional critical boundary in in-plane hBN/graphene interfaces.
	Interface structures and repeat units are shown at left, with the corresponding frequency- and incidence-angle-resolved transmission \(T\) at right.
	(a) Armchair interface.
	(b),(c) Sawtooth interfaces of increasing period \(\delta\).
	The onset of transmission in the conventionally forbidden sector shifts to lower frequency as \(\delta\) increases.
    }
	\label{fig:inplane_transmission}
\end{figure}

\emph{Finite transmission beyond the critical boundary---}We first test whether a periodic crystalline interface can reopen finite transmission in angular sectors that are nominally forbidden in the continuous-interface picture. To this end, we compute frequency- and angle-resolved phonon transmission across in-plane hBN/graphene heterostructures using mode-resolved nonequilibrium Green's functions~\cite{ong2018atomistic,chen2019valley} with interatomic force constants from the neuroevolution potential~\cite{liang2025probing,chen2024unified}, with details given in the Supplemental Material~\cite{SuppM}. We focus on the ZA branch of the commensurate armchair interface~\cite{sutter2012interface,liu2013plane,nguyen2019microscopic} in Fig.~\ref{fig:inplane_transmission}(a), since its quadratic low-frequency dispersion lowers the onset for momentum-compensated channels. Transport is along \(x\), and the incidence angle is defined from the group velocity on the hBN side as \(\theta_i=\arcsin(v_y/|v|)\). Throughout this work, the conventional critical boundary is determined from ordinary \(n=0\) tangential-momentum matching using the lead isofrequency contours, rather than extracted from the transmission spectra.

For the armchair interface, Fig.~\ref{fig:inplane_transmission}(a) shows a strongly frequency-dependent critical boundary, reflecting the quadratic ZA dispersion. Below about \(4.7\) THz, this boundary separates transmitting states from a forbidden sector with essentially no transmission. Above this onset, finite-transmission channels appear outside the conventional critical boundary, where a laterally continuous interface would allow only an evanescent transmitted field. Since the direct \(n=0\) channel is closed in this sector, the finite transmission is a signature of a reopened momentum-compensated channel. The onset agrees with Eq.~(\ref{eq:onset}), occurring when the interfacial momentum shift can return the transmitted state to the propagating sector.

The period dependence provides a sharper test of the mechanism. We therefore compare the armchair interface with sawtooth interfaces whose periods are approximately two and three times larger, as shown in Figs.~\ref{fig:inplane_transmission}(b) and \ref{fig:inplane_transmission}(c). As \(\delta\) increases, \(G_\parallel=2\pi/\delta\) decreases, and the onset of forbidden-sector transmission shifts from about \(4.7\) THz to approximately \(1.5\) and \(0.8\) THz. This systematic shift follows directly from the first-order condition \(q_i(\omega_{\rm on}^{(1)})+q_t(\omega_{\rm on}^{(1)})=G_\parallel\). The sawtooth interfaces also show stronger asymmetry between positive and negative incidence angles. This asymmetry reflects spectral weight redistribution among the allowed compensated channels rather than the criterion for negative refraction itself. This interpretation is supported by the Si/Ge ladder/zipper control interfaces in the Supplemental Material, which have the same \(G_\parallel\) and onset but different angular distributions because of their different lateral inversion symmetries~\cite{PhysRevB.109.075404}. The same qualitative trends are reproduced with an optimized Tersoff potential, indicating that the reopening is not specific to the machine-learned potential (Supplementary Fig.~S3). These results establish the first step of the mechanism. Periodic hBN/graphene interfaces reopen propagating channels in angular sectors that would be totally reflected in the conventional picture. The remaining question is whether these reopened channels also reverse their tangential component and therefore become negative-refraction channels.

\begin{figure}[b]
	\includegraphics{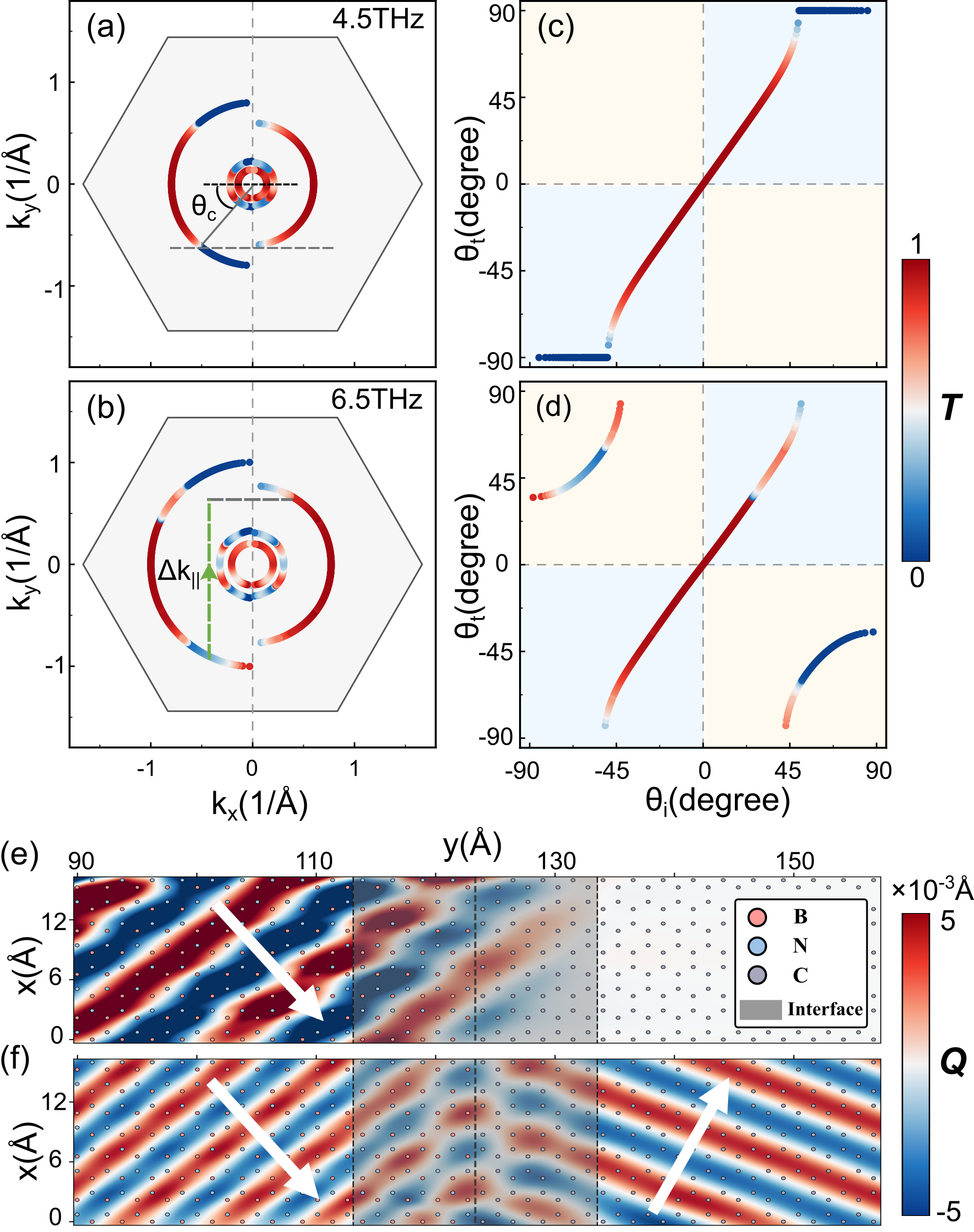}% Here is how to import EPS art
	\caption{
	Momentum-compensated negative-refraction channels at the armchair hBN/graphene interface.
	(a),(b) Unfolded mode-resolved transmission at 4.5 and 6.5 THz, respectively.
	(c),(d) Incidence--transmission angle maps at the same respective frequencies.
	(e),(f) Corresponding transmission eigenchannels beyond \(\theta_c\).
	Here \(\theta_c\) is the conventional critical angle and \(\Delta k_{\parallel}=G_{\parallel}\) is the first-order momentum shift.
	In (e),(f), color and amplitude encode displacement phase and magnitude.
	Only the 6.5 THz case enters the negative-refraction sector and forms a transmitted wavefront.
	}
	\label{fig:negative_refraction}
\end{figure}

\emph{Negative-refraction channels---}We next determine whether the reopened channels identified in Fig.~\ref{fig:inplane_transmission} are negative-refraction channels. We return to the armchair hBN/graphene interface and compare two representative frequencies on opposite sides of the reopening onset. At \(4.5\) THz, the first-order momentum-compensated state remains evanescent, whereas at \(6.5\) THz it has reopened. To resolve the momentum-space character of these channels, we unfold the mode-resolved NEGF transmission spectrum from the lead-supercell Brillouin zone to the primitive Brillouin zone~\cite{boykin2014brillouin}. At \(4.5\) THz, the shift by \(G_\parallel\) is insufficient to bring the transmitted state onto a propagating branch, so the normal wave-vector component remains complex [Fig.~\ref{fig:negative_refraction}(a)]. At \(6.5\) THz, the shifted state lies on a propagating branch and \(k_{t,\parallel}\) reverses sign relative to \(k_{i,\parallel}\) [Fig.~\ref{fig:negative_refraction}(b)]. For the nearly isotropic acoustic phonons considered here, the group velocity is closely aligned with the wave vector, so this sign reversal also reverses the tangential component of the transmitted energy flow. Once the direct \(n=0\) channel is closed, the reopened compensated channel therefore provides an isolated negative-refraction pathway.

This identification is confirmed in incidence-transmission angle space. Figures~\ref{fig:negative_refraction}(c) and \ref{fig:negative_refraction}(d) plot the transmission probability as a function of the incident and transmitted angles. At \(4.5\) THz, transmission remains in the first and third quadrants, corresponding to ordinary positive refraction, and disappears beyond the critical angle \(\theta_c\simeq50^\circ\). At \(6.5\) THz, the reopened channels appear beyond \(\theta_c\) in the second and fourth quadrants, where the incident and transmitted tangential velocity components have opposite signs. This provides the angular-space signature that the momentum-compensated channels in Fig.~\ref{fig:negative_refraction}(b) are negative-refraction channels. A real-space view is provided by the extended phonon transmission-eigenchannel analysis~\cite{klockner2018transmission}. For the \(4.5\) THz case at \(\theta_i\simeq50^\circ\) in Fig.~\ref{fig:negative_refraction}(e), the eigenchannel displacement remains localized near the interface and decays on the transmitted side, as expected for an evanescent wave under total internal reflection. For the \(6.5\) THz case at the same incidence angle, Fig.~\ref{fig:negative_refraction}(f) shows a transmitted wavefront emerging into the graphene side with a reversed tangential propagation component. The eigenchannel therefore visualizes, within the elastic scattering picture, the conversion from total internal reflection to momentum-compensated negative refraction. Together, these momentum-, angle-, and real-space signatures establish the mechanism of momentum-compensated negative refraction in laterally stitched hBN/graphene interfaces. We next use Eqs.~(\ref{eq:window}) and~(\ref{eq:onset}) to design a twist-tunable out-of-plane platform, whose moir\'e superperiod supplies the interfacial reciprocal vector.

\begin{figure}
	\includegraphics{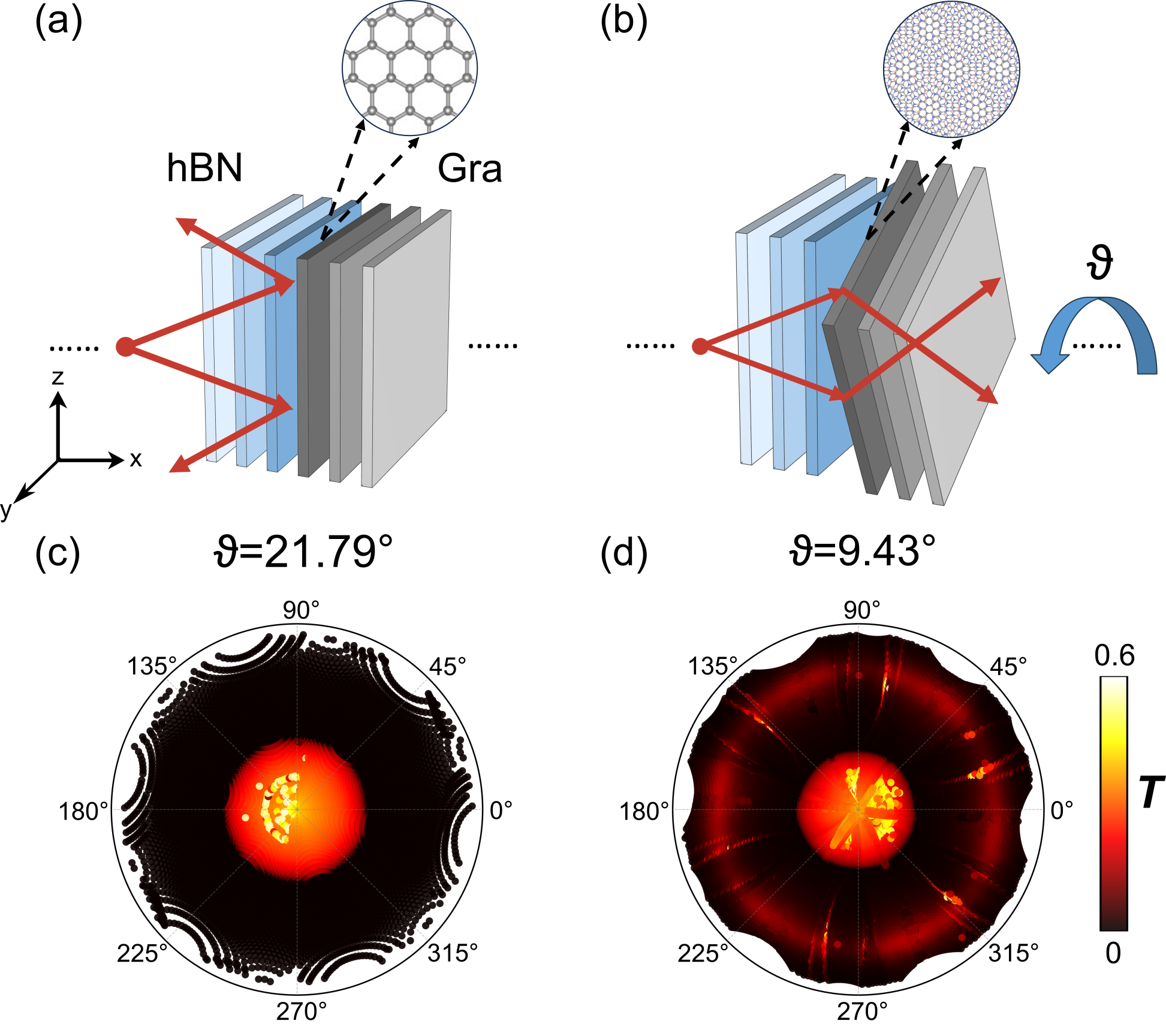}% Here is how to import EPS art
	\caption{
	Twist-controlled opening of negative-refraction channels in out-of-plane hBN/graphene moir\'e heterostructures.
	(a),(b) Schematics of phonon transport through untwisted and twisted structures.
	(c),(d) Angle-resolved transmission \(T\) at 2.5 THz for \(\vartheta=21.79^\circ\) and \(9.43^\circ\), respectively.
	The radial and circumferential coordinates represent \(\theta\), measured from the transport direction, and \(\varphi\), respectively.
	At this frequency, only the \(9.43^\circ\) structure lies inside the first-order momentum-compensation window and opens transmission outside the critical ring.
	}
	\label{fig:moire_transmission}
\end{figure}

\begin{figure}[b]
	\includegraphics{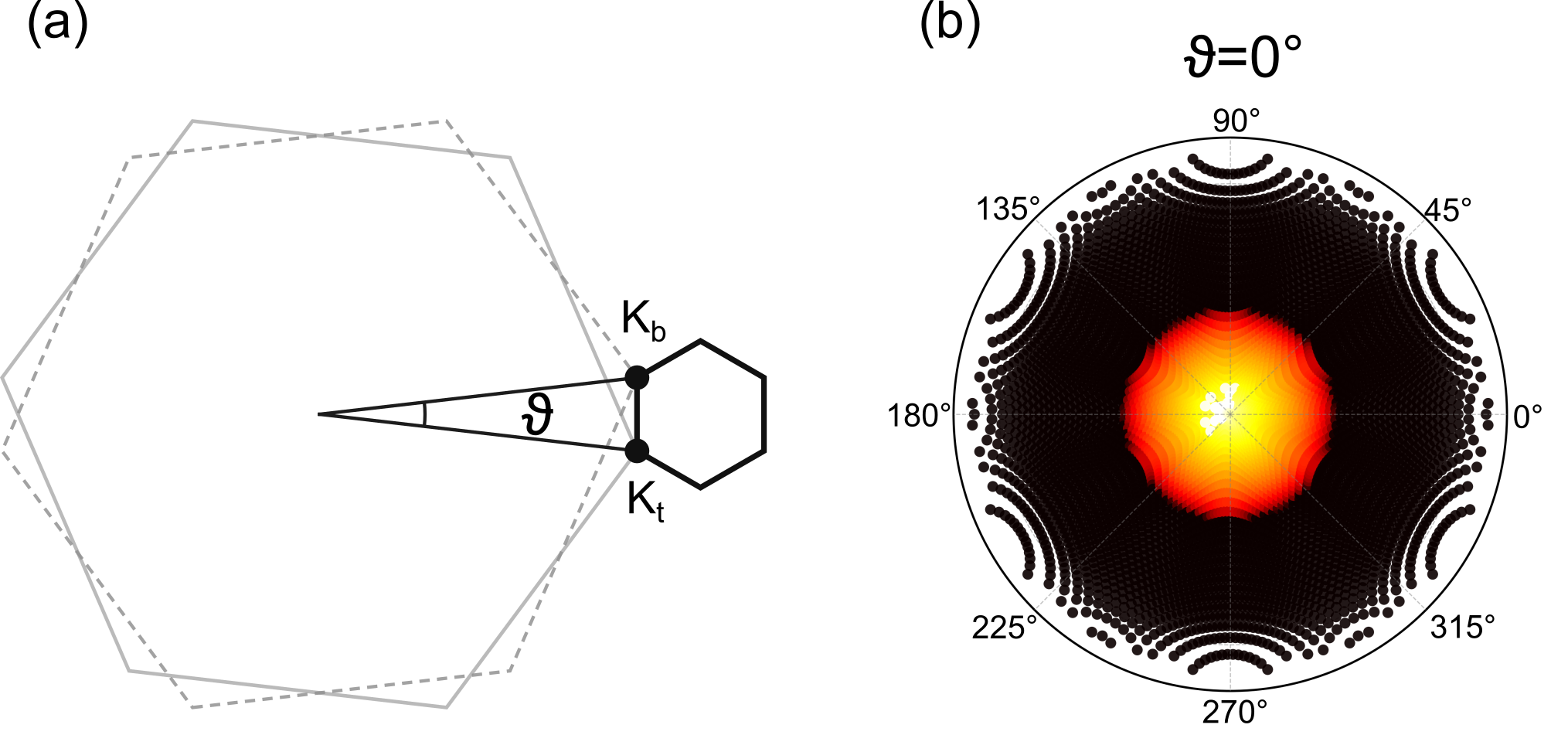}% Here is how to import EPS art
	\caption{
	Reciprocal-space construction and untwisted reference for the moir\'e heterostructure.
	(a) Primitive Brillouin zones of graphene and hBN together with the reduced moir\'e Brillouin zone used to determine the shortest interfacial reciprocal vector \(G_{\rm M}\).
	(b) Angle-resolved transmission \(T\) of the untwisted AA-stacked structure at \(2.5\) THz.
	The radial and circumferential coordinates represent \(\theta\) and \(\varphi\), respectively, and the dashed circle marks the conventional critical angle \(\theta_c\simeq34.4^\circ\).
	}
	\label{fig:moire_reference}
\end{figure}

\emph{Twist-controlled negative refraction---}In twisted out-of-plane hBN/graphene van der Waals heterostructures, the moir\'e superperiod defines an interfacial reciprocal vector \(G_{\rm M}\), whose reciprocal-space construction is shown in Fig.~\ref{fig:moire_reference}(a) of the End Matter. We use Eqs.~(\ref{eq:window}) and~(\ref{eq:onset}) to screen the first-order momentum-compensation window before carrying out full NEGF calculations. This screening determines whether a given twist angle can reopen a propagating state whose tangential momentum is reversed rather than its final angular intensity pattern. For the multilayer hBN/graphene stacks considered here, the relevant out-of-plane acoustic branch is approximately linear over the frequency range of interest. The resulting onset frequencies and upper bounds are summarized in Table~\ref{tab:moire_window} of the End Matter, and a more extensive sequence is given in the Supplemental Material~\cite{SuppM}. As the twist angle decreases, the moir\'e period increases and \(G_{\rm M}\) decreases, shifting the first-order negative-refraction window to lower frequencies. Twist therefore moves the window itself rather than merely changing the transmission strength.

Guided by Table~\ref{tab:moire_window} in the End Matter, we choose \(2.5\) THz because it lies below the first-order onset for \(\vartheta=21.79^\circ\) but within the corresponding window for \(\vartheta=9.43^\circ\). Figures~\ref{fig:moire_transmission}(c) and \ref{fig:moire_transmission}(d) show the respective angle-resolved spectra, while the untwisted AA-stacked reference is given in Fig.~\ref{fig:moire_reference}(b) of the End Matter. Because transport occurs out of plane, the incidence direction is described by polar and azimuthal angles \((\theta,\varphi)\) defined from the group velocity. The \(21.79^\circ\) structure remains below the predicted onset and exhibits negligible transmission beyond the critical ring [Fig.~\ref{fig:moire_transmission}(c)]. By contrast, the \(9.43^\circ\) moir\'e superlattice lies inside the window and supports finite-transmission channels beyond the critical ring [Fig.~\ref{fig:moire_transmission}(d)]. Qualitatively similar behavior occurs for the \(13.17^\circ\) structure (Supplementary Fig.~S5). This controlled contrast verifies that twist switches the compensated negative-refraction channel on or off by moving the operating frequency into or out of the first-order window.

A notable difference from the in-plane hBN/graphene heterostructures is that the finite-transmission ring in the moiré van der Waals system is much more uniform in azimuth. This indicates that the transmitted intensity is redistributed more evenly among channels, consistent with weaker lateral coupling between neighboring moiré units in the out-of-plane geometry. The moiré geometry therefore does more than reproduce the momentum-compensation mechanism. Twist generates the required interfacial reciprocal vector without an atomically stitched covalent boundary, making van der Waals heterostructures a cleaner platform for testing the \(G_{\rm M}\)-controlled on/off behavior. The mechanism's robustness, generality across interface geometries and material classes, and experimental observability are discussed in the End Matter.

\emph{Summary---}We have identified an interfacial momentum-compensation mechanism for THz phonon negative refraction. Discrete translational symmetry along a crystalline interface permits tangential momentum transfer by an interfacial reciprocal vector, allowing transmitted states that would otherwise be evanescent to reopen as propagating channels. When the compensated tangential component reverses sign, these reopened channels become negative-refraction channels even for nearly isotropic acoustic phonons. Mode-resolved calculations for hBN/graphene heterostructures establish this process in laterally stitched in-plane interfaces, where forbidden-sector transmission is reopened and then identified as negative refraction in momentum, angle, and real space. Twisted van der Waals hBN/graphene heterostructures extend the same mechanism to a tunable platform, where the moir\'e superperiod shifts the negative-refraction window to lower frequencies and yields a more azimuthally uniform transmission pattern. These results show that periodic crystalline interfaces can act as symmetry-engineered elements for THz phonon momentum conversion and wavefront control.

\begin{acknowledgments}
\emph{Acknowledgments---}This work was financially supported by the National Key Research and Development Program of Ministry of Science and Technology (Grant No. 2022YFA1402504) and by the National Natural Science Foundation of China (Grants No. 92577103, No. 12374040, and No. 12504258).
\end{acknowledgments}

\nocite{bistritzer2011moire}
\nocite{kresse1996iterative}
\nocite{kresse1996efficiency}
\nocite{togo2008ferroelastic}
\nocite{togo2023phonopy}
\nocite{lindsay2010tersoff}
\nocite{kinaci2012bnc}
\nocite{togo2015lifetimes}
\nocite{liu2023interface_modes}
\nocite{tersoff1989modeling}
\nocite{jiang2019misfit}
\nocite{kandemir2016thermal}
\nocite{mobaraki2018validation}

\bibliography{apssamp}% Produces the bibliography via BibTeX.
%++++++++++++++++++++++++++++++++++++++++++++++++++++++++++++
% ============================================================
%  END MATTER
% ============================================================
%++++++++++++++++++++++++++++++++++++++++++++++++++++++++++++
\onecolumngrid
\vspace{1em}
\begin{center}
	{\large\bfseries End Matter}
\end{center}
\vspace{1em}
\twocolumngrid
%++++++++++++++++++++++++++++++++++++++++++++++++++++++++++++
\emph{Twist-angle screening and reference transmission---}Figure~\ref{fig:moire_reference}(a) shows the reciprocal-space construction used to determine \(G_{\rm M}\). The primitive Brillouin zones of graphene and hBN define the reduced moir\'e Brillouin zone, and \(G_{\rm M}\) denotes the magnitude of the shortest first-order moir\'e reciprocal vector used in the screening. Applying Eqs.~(\ref{eq:window}) and~(\ref{eq:onset}) to the approximately linear out-of-plane acoustic branch gives the twist-angle-dependent windows summarized in Table~\ref{tab:moire_window}. At \(2.5\) THz, the operating frequency lies below the first-order onset for \(\vartheta=21.79^\circ\) but inside the corresponding window for \(\vartheta=9.43^\circ\). The untwisted AA-stacked structure exhibits a well-defined critical ring at \(\theta\simeq34.4^\circ\), with negligible transmission beyond the conventional critical condition [Fig.~\ref{fig:moire_reference}(b)]. This spectrum provides the \(n=0\) reference for the twisted results in Figs.~\ref{fig:moire_transmission}(c) and \ref{fig:moire_transmission}(d) and shows that the channels opened by the \(9.43^\circ\) structure are not a generic feature of out-of-plane transmission.

\begin{table}[b]
	\caption{
		First-order momentum-compensation windows in hBN/graphene moir\'e heterostructures. \(G_{\rm M}\) denotes the magnitude of the shortest first-order moiré reciprocal vectors used for the analytical screening, \(f_{\rm on}^{(1)}\) is the onset frequency, and \(f_{\rm off}^{(1)}\) is the upper edge of the first-order negative-refraction window.
	}
	\label{tab:moire_window}
	\begin{ruledtabular}
		\begin{tabular}{cccc}
			Twist angle & \(G_{\rm M}\) & \(f_{\rm on}^{(1)}\) & \(f_{\rm off}^{(1)}\) \\
			(deg) & (\AA\(^{-1}\)) & (THz) & (THz) \\
			\hline
			21.79 & 0.958 & 3.05 & 6.96 \\
			13.17 & 0.581 & 1.85 & 4.23 \\
			9.43  & 0.417 & 1.33 & 3.03 \\
			7.34  & 0.324 & 1.03 & 2.36 \\
			6.01  & 0.266 & 0.85 & 1.93 \\
		\end{tabular}
	\end{ruledtabular}
\end{table}

\emph{Robustness, generality, and experimental observability---}Our calculations are harmonic and mode resolved and therefore identify elastic interfacial scattering channels selected by interface symmetry. Anharmonic scattering limits the lifetime and directional persistence of the refracted phonon population but does not alter the modulo-\(G_\parallel\) selection rule. The mechanism requires a sufficiently well-defined tangential Fourier component over the coherence length of the incident phonon wave packet rather than an atomically perfect interface. Periodic reconstructions and moir\'e superlattices can meet this condition. The reconstructed 558 hBN/graphene boundary retains momentum-compensated reopening when its tangential periodicity remains well defined (Supplementary Fig.~S6). Branch-resolved mean free paths and \(N={\rm MFP}/\lambda\) support the long-range wave character of the low-frequency acoustic modes, whereas a representative \(22\) THz ZO mode no longer forms a clear beam-like refracted wavefront even when the kinematic condition is satisfied (Supplementary Figs.~S7 and S8). Calculations for zigzag-based hBN/graphene interfaces, Si/Ge interfaces, and TMD in-plane interfaces further show that the \(G_\parallel\)-controlled on/off rule is not specific to hBN/graphene (Supplementary Table~S2 and Supplementary Figs.~S9--S17).

Direct imaging of the mode-resolved eigenchannels in Fig.~\ref{fig:negative_refraction} would require mode-selective excitation and phase-resolved tracking of a single scattering state. A more accessible test is differential. For the same coherent THz phonon wave-packet spectrum, an untwisted stack or a twisted stack below onset should show no focused transmitted-energy pattern beyond the conventional critical condition, whereas a twist angle such as \(9.43^\circ\), whose center frequency lies inside the predicted window, should generate a convergent transmitted-energy pattern. These cases are represented schematically in Figs.~\ref{fig:moire_transmission}(a) and \ref{fig:moire_transmission}(b), respectively. Because the controlled variable is the twist-selected interfacial reciprocal vector, this comparison tests the \(G_{\rm M}\)-controlled momentum-compensation window rather than a generic change in transmission. The key signature is a frequency- and twist-selective redistribution of phonon energy, phonon population, or transient temperature. Van der Waals phononic spectroscopy~\cite{yoon2024terahertz}, spatially resolved vibrational electron energy-loss spectroscopy~\cite{liu2025probing}, and momentum-selective electron energy-loss spectroscopy~\cite{yan2025atomic} provide possible routes to probe these nonequilibrium phonon signatures near interfaces.

%++++++++++++++++++++++++++++++++++++++++++++++++++++++++++++
\end{document}